\documentclass[11pt,onecolumn] {IEEEtran}

\usepackage{graphicx,cite}
\usepackage{latexsym}
\usepackage{multirow}

\usepackage{cite}
\usepackage[cmex10]{amsmath}
\usepackage{bbm}

\usepackage{algorithmic}
\usepackage{array}
\usepackage{mdwmath}
\usepackage{mdwtab}
\usepackage{eqparbox}
\usepackage[tight,footnotesize]{subfigure}
\usepackage[font=footnotesize]{subfig}
\usepackage{fixltx2e}
\usepackage{stfloats}
\usepackage{url}
\begin{document}
\title{Peak Reduction and Clipping Mitigation by Compressive Sensing}

\author{Ebrahim~Al-Safadi and
        Tareq~Al-Naffouri
        % <-this % stops a space
\thanks{The authors are with the Department
of Electrical Engineering, King Fahd University of Petroleum $\&$
Minerals, Dhahran, KSA e-mail:alsafadi@kfupm.edu.sa;naffouri@kfupm.edu.sa}% <-this % stops a space
% <-this % stops a space
\thanks{}}

\maketitle

\begin{abstract}
%\boldmath
This work establishes the design, analysis, and fine-tuning of a
Peak-to-Average-Power-Ratio (PAPR) reducing system, based on
compressed sensing at the receiver of a peak-reducing sparse clipper
applied to an OFDM signal at the transmitter. By exploiting the
sparsity of the OFDM signal in the time domain relative to a
pre-defined clipping threshold, the method depends on partially
observing the frequency content of extremely simple sparse clippers
to recover the locations, magnitudes, and phases of the clipped
coefficients of the peak-reduced signal. We claim that in the
absence of optimization algorithms at the transmitter that confine
the frequency support of clippers to a predefined set of
reserved-tones, no other tone-reservation method can reliably
recover the original OFDM signal with such low complexity.

Afterwards we focus on designing different clipping signals that can
embed a priori information regarding the support and phase of the
peak-reducing signal to the receiver, followed by modified
compressive sensing techniques for enhanced recovery. This includes
data-based weighted $\ell_{1}$ minimization for enhanced support
recovery and phase-augmention for homogeneous clippers followed by
Bayesian techniques.

%Finally, we adapt a greedy Bayesian compressive sensing technique to
%our system that exploits probabilistic information to enhance random
%sparse signal recovery from highly incomplete observations.
%Consequently, a Maximum A Posteriori (MAP) criterion is applied over
%a successively reduced search space for a practical balance between
%performance and complexity.

We show that using such techniques for a typical OFDM signal of 256
subcarriers and 20$\%$ reserved tones, the PAPR can be reduced by
approximately 4.5 dB with a significant increase in capacity
compared to a system which uses all its tones for data transmission
and clips to such levels. The design is hence appealing from both
capacity and PAPR reduction aspects.
\end{abstract}

\begin{IEEEkeywords}
PAPR reduction, tone reservation techniques, compressive sensing,
sparse signal estimation.
\end{IEEEkeywords}

% \fi
\IEEEpeerreviewmaketitle

\section{Introduction}
\IEEEPARstart{D}{espite} the introduction of Single Carrier
Frequency Division Multiple Access (SC-FDMA) into current
multicarrier transmission standards, the success of Orthogonal
Frequency Division Multiplexing (OFDM) in high data rate
transmission remains truly remarkable, with no better proof than the
fact that variants of the IEEE 802.16 and IEEE 802.18 standards are
still emerging \cite{OFDM_applications_3,Wimax2}.

%These signals have infiltrated both wireless and
%wireline communication technologies and only seem to be increasing
%their share in the market. Since 1993, Digital Subscriber Line (DSL)
%adopted OFDM, also called discrete multitone
%(DMT)\footnotemark\footnotetext{which is a modification of OFDM that
%enables bit loading \cite{Tellado}}, following successful field
%trials competitions at Bellcore versus equalizer-based systems
%\cite{OFDM_applications_3}. The European Telecommunications
%Standards Institute (ETSI) has adopted OFDM signaling for European
%Digital Audio Broadcast (DAB) and Digital Video Broadcast (DVB)
%since the late nineties
%\cite{OFDM_applications_5,OFDM_applications_6,OFDM_applications_7,OFDM_applications_8,OFDM_applications_1}.

%Furthermore, OFDM has been accepted for the wireless local area
%network standards from IEEE 802.11, High Performance Local Area
%Network type 2 (HIPERLAN/2) and Mobile Multimedia Access
%Communication (MMAC) Systems
%\cite{OFDM_applications_3},\cite{OFDM_applications_1,OFDM_applications_2,OFDM_applications_4}.
%The multiband OFDM standard for ultrawideband was developed in 2003,
%and variants

The main problem with OFDM signalling however lies in the high
temporal peaks relative to the signal mean, portrayed in a parameter
most commonly referred to as Peak-to-Average-Power-Ratio
(PAPR)\footnotemark\footnotetext{Some authors prefer using ``PAR"
instead for its simpler pronunciation. The fact remains however,
that the problem is in the high frequency power amplifiers and hence
the ratio of powers is the main concern in general.}. Since an OFDM
signal is typically constructed by the superposition of a large
number of modulated subcarriers, its envelope fluctuates with
significant variance, causing the high PAPR. This enforces the use
of expensive Power Amplifiers that should operate linearly over a
wide range of signal amplitudes, which also dissipate a lot of
energy as well \cite{PAPR_overview3}.

Due to the monotonically increasing importance of OFDM signals, the
problem of high PAPR has received considerable attention ever since
OFDM was adopted in important communication standards (see
\cite{PAPR_overview1,PAPR_overview2} for an overview). In the last
decade, the problem of high PAPR in OFDM systems has been tackled by
a variety of approaches, including coding techniques
\cite{coding1,coding3,coding4}, selective mapping
\cite{selected_mapping,selected_mapping_2_low_complexity}, partial
transmit sequences \cite{pts_1,pts_2}, constellation expansion (also
known as tone injection)
\cite{Constellation_reshaping,constellation1,constellation2,Tellado2},
tone-reservation \cite{Tellado,active_set,Tone_reservation_new}, and
companding \cite{companding2,companding4,exponential_companding} to
name a few. Although many of these reduction techniques are
brilliant and very effective, the main obstacle limiting the
implementation of most of them is commonly related to high
complexity \cite{PAPR_overview3}.

In this paper we design, fine-tune, implement, and analyze a novel
tone-reservation based PAPR reducing system that makes a radically
different utilization of these tones compared to previous
techniques. Such a utilization could not have been practically
developed without the implementation of algorithms capable of robust
reconstruction from partial frequency observations. Furthermore, the
application we propose completely switches the stage at which signal
processing complexity is required from the transmitter's side to the
receiver's side of the communication system, and hence provides an
alternate solution to different communication models where the
transmitter's complexity is a bottleneck.

We wish to establish that to the best of our knowledge this is the
first work in the literature where PAPR reduction is achieved using
compressive sensing (CS) \cite{Safadi}. The methods throughout will
always assume sparsity of clipping events relative to a clipping
threshold, and use null tones to estimate these events, providing
the first application of the major work of Candes and Tao on
recovering sparse signals from highly incomplete frequency
information \cite{Candes1} in this context. As such, we also remove
the obstacle faced by all previous tone-reservation-based PAPR
reduction techniques beginning with the pioneering work of Tellado
\cite{Tellado,Tellado2} till very recently
\cite{Chen_tone,Kashin,Shao,Janaaththanan}, all of which required
careful construction of peak-reducing signals at the transmitter in
order to keep them orthogonal to the data signal in the frequency
domain.

Afterwards, we branch off to many solutions to enhance the basic
algorithm by designing different clipping techniques at the
transmitter, modifying the CS algorithm to make use of a priori
support and phase information, and pursuing Bayesian Estimation
techniques for joint support and amplitude estimation at the final
stage.

Unless mentioned otherwise, we use lower case letters for (column)
vectors and upper case letters for matrices. Since we will be
toggling extensively between the time domain and frequency domain,
we will denote by $\check{x}$ the Discrete Fourier Transform (DFT)
of $x$, while we reserve the hat notation $\hat{x}$ to denote the
estimate of $x$. We use $x(i)$ to denote a scalar which is the
$i^{th}$ coefficient of the vector $x$, while we reserve the
subindex notation in $x_{i}$ to denote a vector that is the $i^{th}$
column of the matrix $X$. Furthermore, we denote by $x^{H}$ the
Hermitian conjugate of $x$.

The vectors we treat throughout are complex in general and of
dimension $N$. We denote by $\|x\|_{p}=\left(\sum_{i=1}^{N}
|x(i)|^{p}\right)^{1/p}$ the $\ell_{p}$-norm of a vector $x$ where
$p$ could be an integer or a real number between zero and one. In
the special case where $p=0$ the definition is modified to the
pseudo-norm  $\|x\|_{0}=\sum_{i=1}^{N} q(i)$, where $q(i)=\{1$ if
$x(i)\neq 0$, and $0$ otherwise$\}$.

Although we use the upper case letter $\textbf{F}$ for the Fourier
matrix, it will be clear from context when we also use it to denote
the Cumulative Distribution Function (CDF) of a random variable
$\textbf{x}$, $\mathbbm{F}_{\textbf{x}}(x)$ and Complementary CDF,
$\bar{\mathbbm{F}}_{\textbf{x}}(x)=1-\mathbbm{F}_{\textbf{x}}(x)$.
The Probability Density Function (PDF) will then be denoted by
$f_{\textbf{x}}(x)$. We use $E[\textbf{x}^{m}]$ to denote the
$m^{th}$ central moment of a random variable $\textbf{x}$.

\section{Transceiver Model}

We define the time-domain complex base-band transceiver model as
\begin{equation}
y(k)= \sum_{\ell=0}^{L-1} h(\ell)x(k-\ell)+z(k),
\end{equation}
%\begin{equation}
%y_{k}= \sum_{i=0}^{L-1} h_{l}x_{k-l}+z_{k}
%\end{equation}
where  $\{x(k)\}$ and $\{y(k)\}$ denote the channel scalar input and
output, $h=(h_{0},h_{1},\ldots,h_{L-1})$ is the impulse response of
the channel, $z(k)\sim\mathcal{CN}(0,\sigma_{z}^{2})$ is AWGN. In
matrix form this becomes
\begin{equation}
y=\textbf{H}x+z,
\end{equation}
where $y$ and $x$ are the time-domain OFDM receive and transmit
signal blocks (after cyclic prefix removal) and
$z\sim\mathcal{CN}(\textbf{0},\sigma_{z}^{2}\textbf{I})$.

By the cyclic prefix, $\textbf{H}$ is a circulant matrix describing
the cyclic convolution of the channel impulse response with the
block $x$ and can be decomposed into
$\textbf{H}=\textbf{F}^{H}\textbf{D}\textbf{F}$ where $\textbf{F}$
denotes a unitary Discrete Fourier Transform (DFT) matrix with
$(k,l)$ element
\begin{equation*}
\left[F(k,\ell)\right]=N^{-1/2}\, e^{-j2\pi k\ell/N},\quad k,\ell\in
{0,1,\ldots,N-1}
\end{equation*}
$\textbf{D}=\textmd{diag}(\check{h})$, and
$\check{h}=\sqrt{N}\textbf{F}h$ is the DFT of the channel impulse
response.

\section{Basic PAPR Reduction Design}

\label{basic design}

The time-domain OFDM signal $x$ is typically constructed by taking
the IDFT of the data vector $\check{d}$ whose entries are drawn from
a generic constellation. Since this signal is of high PAPR, we add a
peak-reducing signal $c$ of arbitrary spectral support at the
transmitter and then estimate it and subtract it from the
demodulated signal at the receiver.\

In what follows, the main condition we impose on $c$ is that it be
sparse in time. This is basically the case if we set a clipping
threshold $\gamma$ on the envelope of the OFDM symbols, or if the
transmitter were to clip the highest $s$ peaks. By the incoherence
property of the time-frequency bases \cite{Candes1}, this
necessarily implies that $c$ is then dense (i.e. non-sparse) in the
frequency domain \cite{Tropp4} and such a condition thus cannot be
satisfied in methods where the data and peak-reducing signal must
occupy disjoint tones
\cite{Tellado,Tone_reservation_new,active_set,Chen_tone,Kashin,Shao,Janaaththanan}.
We will denote by $\mathcal{I}_{c}=\{i:\|c(i)\|\neq 0\}$ the sparse
temporal support of $c$ where $|\mathcal{I}_{c}|=s=\|c\|_{0}$.\

%Practically speaking, clipping is done on the oversampled OFDM
%signal to ovoid the problem of peak-regrowth. However, for
%simplicity we assume clipping is done on the OFDM signal sampled at

Throughout this work, we will only consider clipping the Nyquist
rate samples of the OFDM signal. Such a restriction is unnecessary
as it is irrelevant to the data-augmented CS methods we prescribe,
but will otherwise require more elaborate tools such as recent
findings that deal with block sparsity
\cite{Block_Sparsity1,Block_Sparsity2}, and we are forced to delay
such topics for lack of space. With this in mind, following
\cite{Imai} and \cite{Wei} we assume the entries of $x$ will be
uncorrelated and that the real and imaginary parts of $x$ are
asymptotically Gaussian processes for large $N$. This directly
implies that the entries of $x$ are independent and that the
envelope of $x$ can be modeled as a sequence of $iid$ Rayleigh
random variables with a common CDF $\mathbbm{F}_{|X|}(|x|)$ and
parameter $\sigma_{|X|}$ which we will use extensively throughout.

%In future work we address the oversampled
%case and modify the design in this paper to adapt to a more
%practical setting. Although our technique would require more
%reserved frequencies to accommodate the increased nonzero
%coefficients in the oversampled signal $c_{LN\times 1}$ with
%oversampling factor $L$, new tools at our disposal such as
%block-sparsity recovery, correlation information utilization, and
%the increased redundancy in the system can jointly counteract this
%difficulty.

Denoting $\Omega$ as the set of frequencies in an OFDM signal of
cardinality $N$, let $\Omega_{d}\subset\Omega$  be the set of
frequencies that are used for data transmission and
$\Omega_{m}=\Omega\backslash\Omega_{d}$ the complementary set
reserved for measurement tones of cardinality $|\Omega_{m}|=m$. Note
that for compressive sensing purposes, a near optimal strategy is to
use a random assignment of tones for estimating $c$ \cite{Candes2}.
\footnote{Based on results in \cite{Xia} it was found in
\cite{Safadi} and \cite{Naffouri} that by using \textit{difference
sets}, one is able to boost the performance of the recovery
algorithm and reduce the symbol error rate.}

The data symbols $\check{d_{i}}$ are drawn from a QAM constellation
of size $M$ and are supported by $\Omega_{d}$  of cardinality
$|\Omega_{d}|=N-m=k$. Consequently, the transmitted peak-reduced
time-domain signal is
\begin{equation}
\bar{x}=x+c=\textbf{F}^{H}\textbf{S}_{x}\check{d}+c
\end{equation}
where $\textbf{S}_{x}$ is an $N\times k$ selection matrix containing
only one element equal to 1 per column, and with $m$ zero rows. The
columns of $\textbf{S}_{x}$ index the subcarriers that are used for
data transmission in the OFDM system. Similarly, we denote by
$\textbf{S}_{m}$ the $N\times m$ matrix with a single element equal
to 1 per column, that span the orthogonal complement of the columns
of $\textbf{S}_{x}$.\

\noindent Demodulation amounts to computing the DFT
\begin{eqnarray}
\nonumber\check{y}&=&\textbf{F}y=\textbf{F}(\textbf{H}\bar{x}+z)\\
\nonumber&=&\textbf{F}(\textbf{F}^{H}\textbf{DF}(\textbf{F}^{H}\textbf{S}_{x}\check{d}+c)+z)\\
&=&\textbf{D}\textbf{S}_{x}\check{d}+\textbf{D}\textbf{F}c+\check{z}
\end{eqnarray}
where $\check{z}=\textbf{F}z$  has the same distribution of $z$
since $\textbf{F}$ is unitary. Assuming the channel is known at the
receiver, we can now estimate $c$ by projecting $\check{y}$  onto
the orthogonal complement of the signal subspace leaving us with
\begin{eqnarray}
\nonumber \acute{y}&=&\textbf{S}_{m}^{T}\check{y}\\
\nonumber &=&\textbf{S}_{m}^{T}\textbf{DF}c+\acute{z}\\
&=&\Psi c+\acute{z}.\label{y_dash}
\end{eqnarray}
Note that $\acute{z}=\textbf{S}_{m}^{T}\textbf{F}z$ is an $m \times
1$ $i.i.d$ Gaussian vector with a covariance matrix
$\textbf{R}_{\acute{z}}=\sigma_{z}^{2}\textbf{I}_{m\times m}$.

The observation vector $\acute{y}$ is a projection of the sparse
$N$-dimensional peak-reducing  signal $c$ onto a basis of dimension
$m \ll N$ corrupted by $\acute{z}$. To demonstrate how such an
$N$-dimensional vector can be estimated from $m$ linear
measurements, we refer the reader to
\cite{Candes1,Candes2,Donoho2,Tropp2,Tropp3,Fletcher,Wainwright1,Wainwright2},
which also investigate theoretical bounds on $m$, $s$, and $N$ for
guaranteed recovery under various conditions. Note that in our case,
the number of measurements $m$ is equivalent to the number of
reserved tones, while the number of clipped coefficients is
equivalent to $s$, and hence the amount of clipping should be below
certain bounds for reliable recovery given a fixed number of tones
$m$. However, these generic CS bounds will be significantly relaxed
to our advantage in the second part of the paper when we exploit
background information from the data vector $x$.

Now coming back to our problem, assume the peak reducing signal $c$
is $s$-sparse in time, given $\acute{y}$ in (\ref{y_dash}), we can
use any compressive sensing technique at the receiver to estimate
$c$. We will follow the main stream CS literature and use a convex
relaxation of an otherwise NP-hard problem \cite{Tropp3} such as
\begin{eqnarray}
\min_{c\in \textbf{C}^{N}} \|\acute{y}-\Psi c\|_{p}^{p} + \lambda
\|\,c\,\|_{1} \label{CS}
\end{eqnarray}
\noindent for recovery, where $p$ is either $1$ (for basis pursuit
\cite{Donoho1}) or $2$ (for LASSO \cite{Tibshirani}) and $\lambda$
is a parameter for adjusting the sparsity penalty. The resulting
solution by compressive sensing alone is an estimate $\hat{c}_{cs}$
of the peak reducing signal which not only reliably detects the
positions of its nonzero entries, but also gives a good
approximation to the corresponding amplitudes. Notice however that
the estimation of $c$ is by no means restricted to convex
relaxations such as (\ref{CS}), and any compressive sensing method
is valid in general, thus opening the door for many possible
improvements in regard to complexity and efficiency.

\begin{figure}%[htb]
\centering
\includegraphics[width=3.5in]{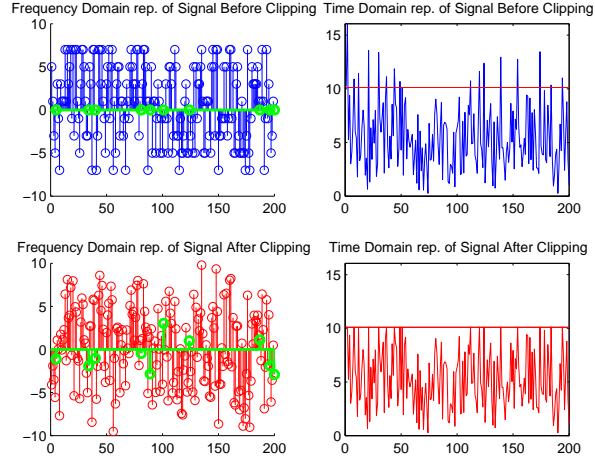}
\caption{Clipping and Tone Reservation} \label{primary_illustration}
\end{figure}

%\begin{figure}[!t]
%\centering
%\includegraphics[width=2.5in]{primary_illustration}
%\caption{Simulation Results} \label{fig_sim}
%\end{figure}

Fig. \ref{primary_illustration} illustrates the main points we've
described so far, although caution must be taken as the actual OFDM
signal is generally complex. %Column-wise, the left column is for the
%frequency domain representation while the right column pertains to
%the time domain representation of the OFDM signal. Row-wise, the
%upper row corresponds to $x$, i.e. the signal before clipping which
%we portray in blue, while the lower row corresponds to $\bar{x}$
%which we portray in red.

\begin{figure}%[htb]
\centering
\includegraphics[width=2.5in]{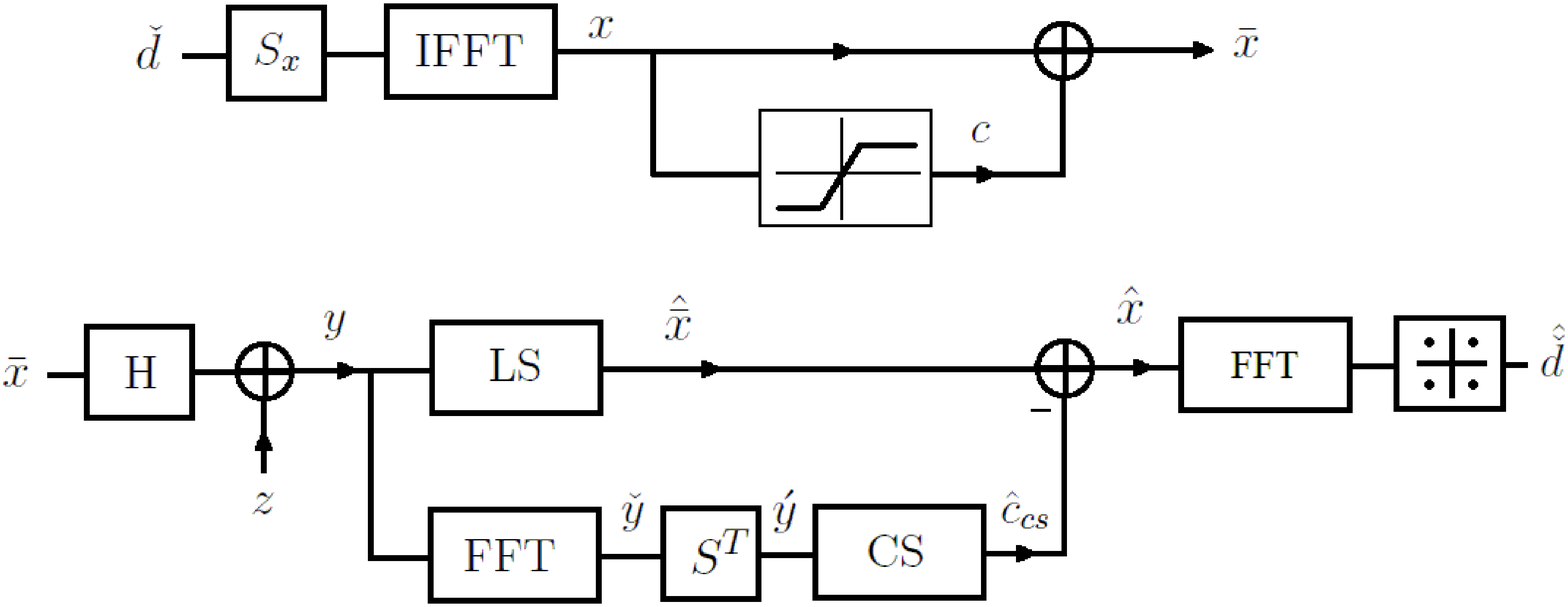}
\caption{Block Diagram of Basic Design} \label{Block Diagram}
\end{figure}

%To describe the sequence of operations at the transmitter, we
%proceed clock-wise through Fig. \ref{primary_illustration} beginning
%with the upper left corner.
%\begin{enumerate}
%\item The frequency domain representation
%$\check{x}$ of $x$ where each coefficient
%$\left\{\check{x}(i)\right\}_{i\in \Omega_{d}}$ in $\check{d}$ is
%drawn from an $M$-QAM constellation. The reserved tones (portrayed
%in green) are now null.
%\item The second figure shows the corresponding time domain representation of the
%OFDM signal, IDFT$(\check{x})$, which is of high PAPR.
%\item The bottom-right figure shows the simple
%operation of clipping the peaks in $x$ and suppressing them to a
%threshold $\gamma$, leaving us with $\bar{x}$ of lower PAPR to
%transmit with the clipping signal $c$ being the difference between
%$x$ and $\bar{x}$.
%\item The last sub-figure shows the effect of clipping in the
%frequency domain. Obviously the distortion has spread over all
%subcarriers including the reserved tones which are no longer null.
%This entire distortion is actually the frequency content of the
%clipping signal $c$ in which we can only partially observe from the
%reserved tones $\Omega_{c}$ and wish to reconstruct $c$ from at the
%receiver.
%\end{enumerate}
The block diagram in Fig. \ref{Block Diagram} stresses that upon
observing $y$, the receiver is confronted with two estimation
problems, the first is the typical estimation of the transmitted
(clipped) OFDM signal $\bar{x}$, and the second is the estimation of
the peak reducing signal $c$. Although the noise statistics are the
same in both cases, the estimation SNR is nevertheless very
different, depending on the clipping procedure. We will hence
reserve the SNR notation for the received signal-to-noise-ratio and
denote by CNR the clipper-to-noise-ratio which is defined as
\begin{eqnarray}
\nonumber \textmd{CNR}&=&\frac{E\left[\|\Psi c\|^{2}\right]}{E\left[\|\acute{z}\|^{2}\right]}\\
&=& \frac{E\left[\|\sum_{k\in
\mathcal{I}_{c}}c(k)\psi_{k}\|^{2}\right]}{\sigma_{z}^{2}}
\end{eqnarray}
and hence depends on the sparsity level
$\|c\|_{0}=|\mathcal{I}_{c}|$ and the magnitudes of $\{c(k)\}_{k\in
\mathcal{I}_{c}}$ which are both functions of the clipping threshold
$\gamma$. This is the parameter of concern when it comes to
compressive sensing in this paper. By definition, the CNR is
typically less than the SNR since the energy of $c$ leaks onto all
the subcarriers even though the CS algorithm only has access to
$\frac{m}{N}$ of them, and also since the magnitudes of the nonzero
coefficients of $c$ are practically smaller than those of $x$.

Note that in using CS our objective is to find the support
$\mathcal{I}_{c}$ of the sparse signal and its complex coefficients
$\{v(k)\}_{k\in\mathcal{I}_{c}}$ at those locations. We could hence
decompose the two problems into $c=\textbf{S}_{c}v_{c}$ and use CS
for the first problem only, giving us $\hat{\textbf{S}}^{(cs)}_{c}$
based on $\hat{\mathcal{I}}^{(cs)}_{c}$, then refine our coefficient
estimate by a more robust technique such as lease squares after
conditioning on the detected support. To do so we define the
$m\times s$ matrix $\hat{\Phi}=\Psi \hat{S}^{(cs)}_{c}$ and refine
our amplitude estimate to
\begin{equation}
\hat{v}^{(ls|cs)}_{c}=(\hat{\Phi}^{H}\hat{\Phi})^{-1}\hat{\Phi}^{H}\acute{y}
\label{ls cs}
\end{equation}
in which
$\hat{c}^{(cs,ls)}=\hat{\textbf{S}}^{(cs)}_{c}\hat{v}^{(ls|cs)}_{c}$
follows. This dual approach is necessary in order to approach an
oracle receiver that uses least squares (see the interesting
discussion in \cite{Wainwright1}).

\section{Comparison with Typical Tone-Reservation PAPR Reduction
Techniques}

The common function of reserved tones in the literature is to act as
a frequency support for the peak reducing signal that is disjoint
from the data-carrying tones
\cite{Tellado,active_set,Tone_reservation_new,Shao,Kashin,Janaaththanan}.
In other words, for each OFDM signal a search is conducted for some
signal $c$ that will reduce the PAPR while being spectrally confined
to a limited number of tones such that
$\|\check{c}\|_{2}-\|\textbf{S}_{m}^{T}\check{c}\|_{2}=0$ and hence
$\check{c}^{\,H}\check{d}=0$. Although many different methods exist
to find such a signal, we only mention the well-known work of
Tellado's \cite{Tellado} for brevity, which requires solving the
convex optimization problem
\begin{eqnarray}
\nonumber &&\min_{\check{c}}\quad t\\
&&s.t.\:\|x+\textbf{F}^{H}\textbf{S}\check{c}\|^{2}\leq t
\label{Tellado}
\end{eqnarray}
where $\check{c}=\textbf{F}c$ is nonzero only on $\Omega_{c}$ from
the definition of $\textbf{S}$. Clearly, this optimization approach
should result in significantly more PAPR reduction compared to our
design, since for the same number of reserved tones $m$, we can only
clip $s < m$ maximum peaks, whereas by Tellado's method no such
restriction exists.

Most importantly however, the main complexity (i.e. the stage at
which the optimization search is performed) in these techniques is
at the transmitter, since the main concern is to find $c$ that will
reduce the PAPR while occupying completely disjoint tones in order
to remain discernable at the receiver.

%It is worth mentioning that Tellado also suggests a suboptimal
%system that avoids repetitive optimization by designing \textit{peak
%reducing kernels}. These are used for directly clipping the peaks
%which exceed a threshold (similar to our basic threshold-based
%clipper), but are designed at initialization to have a confined
%spectral support based on the provided peak reducing tones. The
%problem is that due to this confined spectral support, these kernels
%have temporal sidelobes that cause variable peak-regrowth among the
%OFDM signals, and Tellado attempts to optimize the location of the
%peak reducing tones to minimize these sidelobes. Clearly, our
%approach sidesteps this annoying problem since we need not confine
%the spectral support of $c$.\

\section{Enhanced PAPR Reduction by Data-Induced Weighted and
Phase-Augmented $\ell_{1}$ Minimization} \label{WCS}

So far we were only interested in using compressive sensing in its
most abstract form as it applies to our problem. We assumed,
following the general literature on CS, that absolutely no
information is known about the locations, magnitudes, and phases of
the sparse signal $c$, beyond the incomplete frequency observations
which we obtained from the reserved tones $\Omega_{c}$
\cite{Candes1,Candes2}. In other words, the model $\acute{y}=\Psi
c+\acute{z}$ was assumed to exist independently of the general
transceiver model $y=H\bar{x}+z$, even though in reality we know
that $c$ is intimately linked to $\bar{x}$ by the simple fact that
it's superimposed on $x$ in the time domain.

%This implies that, although we cannot directly make use of the
%frequency content of the other subcarriers $\Omega_{d}$ since there
%is no way to distinguish between the portions of each signal $x$ and
%$c$ on those frequencies, we can still obtain valuable information
%about $c$ from the \emph{time domain} representation of the
%estimated clipped OFDM signal $\hat{\bar{x}}$.

The upshot of this section is to demonstrate that for optimal PAPR
reduction using CS, the estimation of the clipping signal at the
receiver should exploit as much information as possible in both
basis representations, which can be achieved by weighting,
constraining, or rotating the frequency-based CS search, based on
information we infer from the data in the time domain.

The difficulty of these problems is strongly related to the way
clipping is performed. Although we have full control in selecting
the sparsity level and the clipping magnitudes and phases to best
suite our purpose, there can't be a clipping technique that
optimizes both the support recovery \emph{and} coefficient
estimation, and a compromise must be made regarding the quality of
the two.
%In particular, depending on the way $c$ is constructed, we can to
%some extent infer locations of the nonzero coefficients of $c$ from
%certain blueprints in $\hat{\bar{x}}$, instead of naively assuming a
%uniform probability distribution over all sparsity patterns in
%$\textbf{J}$ prior to CS. Likewise, in some cases we know that the
%magnitudes of the nonzero coefficients of $c$ have random magnitudes
%that are conditioned on the magnitudes of $x$. Finally, most
%constructions of $c$ will draw the phases of the nonzero
%coefficients of $c$ directly from the clipped coefficients of $x$.
%We should thus emphasize the following concept which is one of the
%main themes of this work.

%Obviously, exploiting all the available information implies using
%the probability density functions of the random variables we wish to
%estimate.
\subsection{Homogeneous Clipping Techniques}

we first begin with defining two simple clipping techniques that do
not require any optimization or spectral confinement, and although
we derive their PDFs along other properties, we focus exclusively on
deterministic CS enhancement techniques\footnote{Although the LASSO
estimate has a MAP interpretation \cite{Tibshirani} we don't assume
any prior or statistic is used.}, and delay the matter of Bayesian
compressive estimation or sensing to the following section.

\subsubsection{Peak Suppression to $\gamma$ (PS)} \label{PS}

\begin{figure}%[htb]
\centering \includegraphics[width=2.5in]{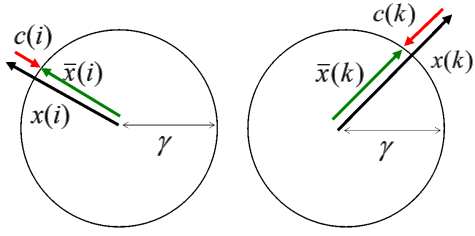}
\caption{Peak Suppression Illustrated on the Complex Plane}
\label{PS figure}
\end{figure}

Because clipping is done on the coefficients of $x$ whose envelope
exceed $\gamma$, the most natural construction of the clipping
signal $c$ would be to basically suppress the magnitudes of the
entries $x_{i}:|x_{i}|\geq\gamma$ to $\gamma$ while preserving their
angles, such that $|x_{i}+c_{i}|=\gamma$ (see Fig. \ref{PS figure}).
This is commonly expressed in the literature
\cite{Bahai,Capacity_clipped_1} as
%\begin{eqnarray}
%\bar{x}_{i}=\left\{\gamma e^{\;j\angle x_{i}}\; if \;|x_{i}|>\gamma,
%\;x_{i}\; otherwise\right\}.
%\end{eqnarray}
\begin{eqnarray}
\bar{x}(i) =
\begin{cases}
\gamma e^{\;j\theta_{x(i)}}\;  & \mbox{if} \;|x(i)|>\gamma, \\
\,\,\,\,x(i)\; & \mbox{otherwise}
\end{cases}
\end{eqnarray}
%Consequently, the clipping signal will take the form
%\begin{eqnarray}
%c^{ps}=-\sum_{i\in
%\mathcal{I}_{c}}\left(|x(i)|-\gamma\right)e^{\;j\theta_{x(i)}}\delta(n-n(i)).
%\end{eqnarray}

%\begin{figure}%[htb]
%\begin{center}
%\epsfxsize = 3.0 true in \epsfbox{twi_sided_pdfs_c_PS.eps}
%\caption{Probability Density Function of $c$ for Peak Suppression as
%a Function of $\gamma$} \label{pdf_c_ps_illustration}
%\end{center} \end{figure}

Obviously, the PDF of the nonzero coefficients of $c^{ps}$ will
depend on the PDF of $|x|\big||x|>\gamma$. Hence if we define the
binary set $\mathcal{Q}$ to label the mutually exclusive events of
clipping or not at a certain index $i$ then
\begin{eqnarray}
\nonumber f(|c^{ps}|(i))&=&\sum_{q\in \mathcal{Q}} P\left(|c^{ps}(i)|\,|q\right)P(q)\\
\nonumber &=& f_{|X|
||X|>\gamma}(|c^{ps}(i)|+\gamma)\left(\bar{\mathbbm{F}}_{|X|}(\gamma)\right)\\
\nonumber & & \quad\quad+\;\delta(|c^{ps}(i)|)\mathbbm{F}_{|X|}(\gamma)\\
\nonumber &=&
\alpha^{-1}(\gamma)\left(\bar{\mathbbm{F}}_{|X|}(\gamma)\right)f_{|X|}(|c^{ps}(i)|+\gamma)\\
& &  \quad\quad \cdot \;
u(|c^{ps}(i)|)+\mathbbm{F}_{|X|}(\gamma)\delta(|c^{ps}(i)|)
\label{pdf_c_ps}
\end{eqnarray}
%\begin{IEEEeqnarraybox}{rCl}
%\nonumber f(|c^{ps}|(i))&=&\sum_{q\in \mathcal{Q}} P\left(|c^{ps}(i)|\,|q\right)P(q)\\
%\nonumber &=& f_{|X|
%||X|>\gamma}(|c^{ps}(i)|+\gamma)\left(\bar{\mathbbm{F}}_{|X|}(\gamma)\right)\\
%\nonumber & & \quad\quad+\;\delta(|c^{ps}(i)|)\mathbbm{F}_{|X|}(\gamma)\\
%\nonumber &=&
%\alpha^{-1}(\gamma)\left(\bar{\mathbbm{F}}_{|X|}(\gamma)\right)f_{|X|}(|c^{ps}(i)|+\gamma)\\
%& &  \quad\quad \cdot \;
%u(|c^{ps}(i)|)+\mathbbm{F}_{|X|}(\gamma)\delta(|c^{ps}(i)|)
%\label{pdf_c_ps}
%\end{IEEEeqnarraybox}
\noindent where $u(\cdot)$ is the unit step function and
$\alpha(\gamma)=\int_{\gamma}^{\infty}f_{|X|}\left(|x|\right)dx$ is
a normalizing constant which depends only on $\gamma$ and is
required to ensure that $\int_{0}^{\infty}
f_{|X|||X|>\gamma}\left(|x|||x|>\gamma\right)d|x|=1$. Not
surprisingly, this is the most popular soft clipping scheme due to
its simplicity and relatively low spectral distortion.

%\noindent Observing from a cross-section on the complex plane,
%Figure \ref{pdf_c_ps_illustration} illustrates the change in a
%hypothetical PDF of a single coefficient of $c^{ps}$ as a function
%of $\gamma$. Notice that for high clipping thresholds (corresponding
%to large values of $\gamma$), the PDF has a sharp peak and a concave
%decay at its sides similar to a Laplacian random variable (albeit
%having a stronger decay as the tail of a Gaussian). Then as we
%decrease $\gamma$, the PDF tends to replicate the PDF of the data
%coefficient $x$ of which it's originally drawn upon, ultimately
%resembling a Gaussian in case $x$ happens to be so as well.

%Since the nonzero coefficients of $c^{ps}$ are $iid$, this can be
%written more compactly by taking the joint PDF of $c$ with support
%set $\textbf{J}$ and then marginalizing over $\textbf{J}$:
%\begin{eqnarray}
%\nonumber f(|c|^{ps})&=&\sum_{i=1}^{2^{N}}P(|c|^{ps}|J_{i})P(J_{i})\\
%&=&\sum_{i=1}^{2^{N}}[f_{|X|||X|>\gamma}(|c^{ps}|+\gamma)]^{|J_{i}|}\;(1-F_{|X|}(\gamma))^{|J_{i}|}(F_{|X|}(\gamma))^{N-|J_{i}|}
%\end{eqnarray}

Two features of this clipping scheme stand out in regard to CS
enhancement. The first is that by suppressing all the data
coefficients to a fixed and known threshold value $\gamma$, we could
actually infer some additional information regarding possible
clipping locations from the distance between the estimated
coefficients' magnitudes and $\gamma$. This clipping scheme can
hence provide additional information regarding the support
$\mathcal{I}_{c}$. The second feature is that the nonzero
coefficients of $c^{ps}$ are exactly anti-phased with the data
coefficients at $\mathcal{I}_{c}$\footnote{we will call such signals
\emph{homogeneous} clippers since their phases are aligned with the
data.}, giving us another source of information regarding the phases
$\theta_{c^{ps}(\mathcal{I}_{c})}$ based on $\hat{\bar{x}}$.

In terms of delectability from standard compressive sensing,
however, the method is quite un-satisfying if left un-enhanced,
demanding a higher number of measurements for the same sparsity
level and Symbol Error Rate (SER) compared to other clipping
techniques. The main reasons are
\begin{enumerate}
\item \textbf{Low CNR}: The CNR in PS decreases very rapidly with $\gamma$. Assuming we neglect
the effect of $\Psi$,
\begin{eqnarray}
\nonumber E[\|c^{ps}\|^{2}]&=& \sum_{k\in \mathcal{I}_{c}}
E\left[|c^{ps}(k)|^{2}|\right]\\
\nonumber&=& E\left[|c^{ps}(k)|^{2}\right]\cdot E[\|c\|_{0}]\\
\nonumber&=& \int_{\infty}^{\infty} |c^{ps}(k)|^{2}f(|c^{ps}(k)|)d|c^{ps}(k)|\\
\nonumber& & {} \quad\quad \cdot E[\|c\|_{0}]\\
&=&
\nonumber\left[\alpha^{-1}(\gamma)(2\sigma_{|X|}^{2}+\gamma^{2})e^{-\frac{\gamma^{2}}{2\sigma_{|X|}^{2}}}-\gamma\right]\\
& & {} \quad\quad \cdot E[\|c\|_{0}]
\end{eqnarray}
where the average sparsity
\begin{eqnarray*}
E[\|c\|_{0}]&=&N^{2}\left(\bar{\mathbbm{F}}_{|X|}(\gamma)\right)^{2}-N\left(\bar{\mathbbm{F}}_{|X|}(\gamma)\right)^{2}\\
& & {} \quad\quad+N\left(\bar{\mathbbm{F}}_{|X|}(\gamma)\right)
\end{eqnarray*}
is simply the expectation of the Binomial corresponding to the
sparsity level. Notice the accumulative effect of $\gamma$ on
$E[\|c^{ps}\|^{2}]$.
\item \textbf{The vanishing of $|c^{ps}_{min}|$}: the random magnitudes of
$c^{ps}$ are drawn from the tail distributions of the data
coefficients, making the limiting distance between the minimum
penetrating coefficient and $\gamma$ approach zero. This is a
critical bottleneck in CS that cannot be completely compensated for
by increasing the CNR. Fletcher et al. \cite{Fletcher} and
Wainwright \cite{Wainwright1,Wainwright2,Wainwright3} stress this
point.
\end{enumerate}

\subsubsection{Digital-Magnitude Clipping (DMC)} \label{dc}
%
%With all disregard to other parameters, the class in which the
%nonzero coefficients of the sparse signal $c$ fall in has a
%significant effect on the quality of the compressive sensing
%results.

\begin{figure}%[htb]
\centering \includegraphics[width=2.5in]{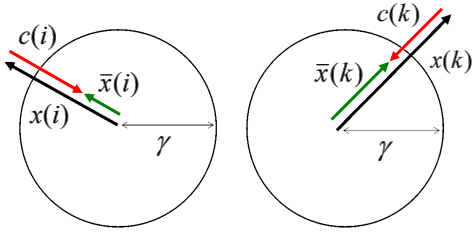}
\caption{Clipping with Fixed Magnitude $\zeta$}
\label{illustration_dc}
\end{figure}

In order to avoid the problems of the previous clipping technique,
we could increment the magnitudes of $c^{ps}$ by some constant until
we're satisfied with the CNR and $|c^{ps}_{min}|$. This however
still leaves us with the burden of estimating the random magnitudes
while destroying the enhanced support detection property of peak
suppression. Instead, consider inverting the procedure from
suppressing \emph{to} a fixed value $\gamma$, to suppressing
\emph{by} a fixed value $\zeta$. \footnote{Quite expectedly, in
\cite{Fletcher} it was shown that, with no modification or
realization to this additional structure, a compressive estimation
algorithm works best when all the nonzero coefficients in $c$ are
equal in magnitude.}

Now that $\{|c(k)|\}_{k\in \mathcal{I}_{c}}=\zeta$, we've decreased
the degrees of freedom of $c$ to $\mathcal{I}_{c}$ and $\theta_{c}$
only. Furthermore, such a clipping scheme preserves the anti-phase
property as well, thus possibly reducing the problem to that of
support detection. \footnote{In the case of digital clipping with
phase augmentation, the problem can also be recast as that of
\emph{detecting} a point on a sparse lattice, and a regularized
sphere decoding algorithm could be used
\cite{Rank_Deficient_Sphere,Finite_Alphabet,Giannakis_Finite_Alphabet}.}

More generally, we could suppress the high peaks of $x$ by a finite
set of magnitudes $\{\zeta_{0},\zeta_{1},\ldots,\zeta_{\ell}\}\in
\mathbbm{Z}^{\ell}$, hence the attribute of Digital Magnitude
Clipping (or simply Digital Clipping for short), although we will
only focus here on
the binary magnitude space $|c|\in\{0,\zeta\}$. %Consequently, the
%nonzero coefficients of $c$ will now lie on a circle of radius
%$\zeta$ in the complex plane, which should not be confused with the
%complex plane wherein the OFDM coefficients $\{x_{i}\}$ lie.

Following the same procedure in finding (\ref{pdf_c_ps}), and by
noting the interesting relation $\|c\|_{p}=\zeta\|c\|_{0}^{^{1/p}},
\,\,p=1,2,..,$ the PDF of the clipping signal's envelope is
basically
\begin{eqnarray}
f\left(|c|^{dm}(i)\right)=\left(\bar{\mathbbm{F}}_{|X|}(\gamma)\right)\delta\left(|c|-\zeta\right)+\mathbbm{F}_{|X|}(\gamma)\delta\left(|c|\right).
\end{eqnarray}
%\begin{eqnarray}
%\nonumber P(|c|^{bm})&=&\prod_{i=1}^{N} P\big(c^{bm}(i)\big)=\prod_{i=1}^{N} \sum_{t=0}^{1}P(c_{i}|\lambda_{t})P(\lambda_{t})\\
%\nonumber &=&\prod_{i=1}^{N} P(c_{i}||x_{i}|\geq \gamma)P(|x_{i}|\geq \gamma)+P(c_{i}||x_{i}|< \gamma)P(|x_{i}|< \gamma)\\
%&=&\prod_{i=1}^{N}
%(1-F_{|X|}(\gamma))\delta(|c|-\zeta)+F_{|X|}(\gamma)\delta(|c|)
%\end{eqnarray}
The PDF of a coefficient's magnitude has been reduced to a Bernoulli
random variable with probability of success
$\left(\bar{\mathbbm{F}}_{|X|}(\gamma)\right)$. Furthermore, the two
clipping methods PS and DMC achieve the same CNR when
\begin{eqnarray}
\zeta=\sqrt{\alpha^{-1}(\gamma)\left(2\sigma_{|X|}^{2}+\gamma^{2}\right)e^{\gamma^{2}/2\sigma_{|X|}^{2}}-\gamma}.\label{CNR_equivalence}
\end{eqnarray}

%\begin{figure}%[htb]
%\begin{center}
%\epsfxsize = 2.0 true in \epsfbox{illustration_dc.eps}
%\caption{Clipping with Fixed Magnitude $\zeta$}
%\label{illustration_dc}
%\end{center} \end{figure}

There is a conflicting interest in deciding the value of $\zeta$. On
one hand, the more we increase it the higher the CNR and the easier
the support detection becomes, but on the other, the overall error
of the system dramatically increases in case of faulty support
detection. Furthermore, oversampling at the subsequent stage of
transmission becomes more complex in this latter case.

Nevertheless, we should at least set a lower bound on its value to
ensure that all clipped coefficients will always end up with
magnitudes equivalent to or bellow the desired clipping threshold
$\gamma$, depending on the envelopes maximum order statistic.
Afterwards, we should be very conservative in increasing $\zeta$

\subsection{Externally Weighted $\ell_{1}$
Minimization} \label{EWCS}

If by some prior information we have a better picture regarding the
support $\mathcal{I}_{c}$ beyond the Bernoulli process assumption,
we can modify the LASSO in (\ref{CS}) by penalizing disfavored
locations so that
\begin{eqnarray}
\hat{c}=\arg\min \|\acute{y}-\Psi c\|_{2}^{2}+\lambda\|w^{T}c\|_{1},
\label{weighted LASSO}
\end{eqnarray}
where $w$ is a weighting vector imposed on the $\ell_{1}$ penalty
term based on this prior information. In the literature, the source
of $w$ is from previous runs of the CS algorithm itself
\cite{Candes4}\cite{Wipf1}, where the hope is that with each
iteration more confidence will exist in
$\hat{\mathcal{I}}^{(k+1)}_{c}$ based on, for instance
\cite{Candes4},
\begin{eqnarray}
w(i)^{(k+1)}\propto
\left[\;|\hat{c}(i)^{(k)^{cs}}|+\epsilon\right]^{-1}\qquad
i=1,2,\ldots,N \label{w_candes}
\end{eqnarray}
where $\epsilon>0$ is a small stabilizing parameter. We will refer
to this procedure as \emph{internally} weighted $\ell_{1}$
minimization.

Repeating the CS algorithm is computationally expensive, and the
process is sensitive to the quality of the first unguided CS
estimate. Instead, we would rather use a one-shot weighting scheme
that minimally increases the complexity of an ordinary LASSO.
Fortunately, this could be done if we had an \emph{external} source
of information based on the data vector $\hat{\bar{x}}$.

Recall the discussion in \ref{PS} regarding embedded information on
the support $\mathcal{I}_{c}$ in peak suppression. The idea is that
we expect the coefficients of $\hat{\bar{x}}$ whose magnitudes are
close to $\gamma$ to be more probable clipping locations compared to
ones that are not. Consequently, we can define a weighting vector
$w^{ps}$ based on the distance
\begin{eqnarray}
d(i)=||\hat{\bar{x}}(i)|-\gamma|,\qquad
i=1,2,\ldots,N\label{ps_distance}
\end{eqnarray}
and use it in (\ref{weighted LASSO}). Another data-based weighting
scheme would be the posterior probability of not having a clip
($q=0$) given the observation (\ref{ps_distance}), such that less
likely clipping locations are more severely penalized by having a
higher such posterior probability
\begin{eqnarray}
w(i)^{ps}&=&\Pr\left(q=0\,|\,d(i)\right)\\
\nonumber &=& \frac{\Pr(d(i)|q=0)\Pr(q=0)}{\sum_{q\in
\mathcal{Q}}\Pr(d(i)|q)\Pr(q)}\\
\nonumber &=&
\frac{f_{|\hat{X}|}(\gamma-d(i))\mathbbm{F}_{|X|}(\gamma)}{f_{|\hat{X}|}(\gamma-d(i))\mathbbm{F}_{|X|}(\gamma)+f_{|E|}(d(i))\bar{\mathbbm{F}}_{|X|}(\gamma)}
\label{ps_distance_prob}
\end{eqnarray}
%\begin{eqnarray}
%w(i)^{ps}&=&\Pr\left(q=1|d(i)\right)\\
%\nonumber &=& \frac{\Pr(d(i)|q=1)\Pr(q=1)}{\sum_{q\in
%\mathcal{Q}}\Pr(d(i)|q)\Pr(q)}\\
%\nonumber &=&
%\frac{f_{|E|}(d(i))\bar{\mathbbm{F}}_{|X|}(\gamma)}{f_{|E|}(d(i))\bar{\mathbbm{F}}_{|X|}(\gamma)+f_{|X|}(\gamma-d(i))\mathbbm{F}_{|X|}(\gamma)}
%\label{ps_distance_prob}
%\end{eqnarray}
where $f_{|E|}$ is the density function corresponding to the
estimation error of the data envelope $|\hat{x}|(i)$, which is the
sole reason $d(i)\!>\!0$ when conditioned on clipping $x(i)$. Using
least squares to recover $x(i)$, we assume its error to be Gaussian
and hence $f_{|E|}$ and $f_{|\hat{X}|}=f_{|X+E|}$ to be Rayleigh
with parameters $\sigma_{|E|}$ and
$\sigma_{|X+E|}=\left[2^{-1}(\sigma_{X}^{2}+\sigma_{E}^{2})\right]^{1/2}$,
respectively. Defining
$\eta(\gamma)=1-e^{-\gamma^{2}/\sigma_{X}^{2}}$, this becomes
%\begin{eqnarray}
%w(i)^{ps}=\frac{\frac{2\eta(\gamma)
%(\gamma-d(i))}{\sigma_{X}^{2}+\sigma_{E}^{2}}e^{-\frac{(\gamma-d(i))^{2}}{\sigma_{X}^{2}+\sigma_{E}^{2}}}}{\frac{2\eta(\gamma)
%(\gamma-d(i))}{\sigma_{X}^{2}+\sigma_{E}^{2}}e^{-\frac{(\gamma-d(i))^{2}}{\sigma_{X}^{2}+\sigma_{E}^{2}}}
%+ \frac{2(1-\eta(\gamma))d(i)}{\sigma_{E}^{2}}
%e^{-\frac{d(i)^{2}}{2\sigma_{E}^{2}}}}\label{ps_distance_prob_final}
%\end{eqnarray}

\begin{eqnarray}
w(i)^{ps}&=&\frac{\diamondsuit e^{\clubsuit}}{\diamondsuit
e^{\clubsuit}+ \triangle e^{\spadesuit}}\label{ps_distance_prob_final}\\
\nonumber &=&\begin{cases} \frac{\diamondsuit}{\diamondsuit+\triangle e^{\spadesuit-\clubsuit}}   ;  & \mbox{if} \;|\clubsuit|>|\spadesuit|, \\
\frac{\diamondsuit e^{\clubsuit-\spadesuit}}{\diamondsuit
e^{\clubsuit-\spadesuit}+\triangle} & \mbox{if}
\;|\spadesuit|>|\clubsuit|
\end{cases}
\end{eqnarray}

%\begin{eqnarray}
%w(i)^{ps}&=&\frac{\diamondsuit e^{\clubsuit}}{\diamondsuit
%e^{\clubsuit}+ \triangle e^{\spadesuit}}\\
%\nonumber &=&\begin{cases} \frac{\diamondsuit}{\diamondsuit+\triangle e^{\spadesuit-\clubsuit}}   ;  & \mbox{if} \;|\clubsuit|>|\spadesuit|, \\
%e^{\log\left(\frac{\diamondsuit}{\triangle+\diamondsuit
%e^{\clubsuit-\spadesuit}}\right)+\clubsuit-\spadesuit}\; & \mbox{if}
%\;|\spadesuit|>|\clubsuit| \label{ps_distance_prob_final}
%\end{cases}
%\end{eqnarray}
where
\begin{eqnarray}
\nonumber \diamondsuit
=\frac{2\eta(\gamma)(\gamma-d(i))}{\sigma_{X}^{2}+\sigma_{E}^{2}}&,&
\clubsuit=\frac{(\gamma-d(i))^{2}}{\sigma_{X}^{2}+\sigma_{E}^{2}}\\
\nonumber \triangle =\frac{2(1-\eta(\gamma))d(i)}{\sigma_{E}^{2}}&,&
\spadesuit=\frac{d(i)^{2}}{\sigma_{E}^{2}}.
\end{eqnarray}
The second part of (\ref{ps_distance_prob_final}) is a necessary
manipulation for numerical stability.

Notice also that what helps in suppressing only to $\gamma$ here is
that we have a probabilistic means to cast out most of the possible
false positives. Had we suppressed the magnitudes to the envelope
mean for instance, $E[|x(i)|]$, the procedure above would favor many
locations as clipping positions by the fact that
$|\hat{\bar{x}}|-E[|x(i)|]$ is small. Nonetheless, misleading bias
to certain locations as candidates for clipping positions due to
their coefficient's natural proximity to $\gamma$ can never be
completely eliminated, even at infinite CNR.

\subsection{Phase-Augmented CS for Homogenous Clippers} \label{Phase
Augmented CS}

In the case of homogenous clipping,
$\theta_{c}(\mathcal{I}_{c})=\theta_{\bar{x}}(\mathcal{I}_{c})$ at
the transmitter, and consequently the CS algorithm should have
access to additional information regarding the phases of the nonzero
coefficients. The problem however is that we only have an estimate
$\theta_{\hat{\bar{x}}}(\mathcal{I}_{c})$ at the receiver, and the
extent to which CS can benefit from this property depends on how
good the estimate $\hat{\bar{x}}$ is in general. To this end, we
will only consider the SNR as the parameter to which we judge the
quality of the data estimate.

Recall the discussion following Fig. \ref{Block Diagram} regarding
the CNR and SNR, and consider the effect of gradually increasing
$\zeta$ which we defined in \ref{dc}. Notice that when $\zeta=0$,
the $\gamma$-penetrating coefficient attains its maximum SNR, then
as we increase $\zeta$ the CNR increases as
$\zeta^{2}E\left[\|c\|_{0}\right]$ while the SNR decreases by
$\zeta\left(2E[|x|]-\zeta\right)$. Consequently, the CNR will be
larger than the SNR in the locations where
$\left(E\left[\|c\|_{0}\right]-1\right)\zeta^{2}+2E\left[|x|
\right]\zeta-E\left[|x|^{2}\right]>0$. Fortunately practical values
of $\zeta$ relative to $E\left[|x|\right]$ fall outside this region,
and we would normally expect to gain information regarding
$\theta_{c}$ from $\hat{\bar{x}}$ that is more reliable than
information from CS alone.

This fact encourages us to absorb, and perhaps even \emph{replace}
altogether, as much information as possible regarding $\theta_{c}$
from the estimated data vector $\hat{\bar{x}}$. Assume first that we
know the vector $\theta_{c}$, we could then merge this information
into the CS algorithm by expressing the clipping signal as
$c=\Theta_{c} |c|$ such that
\begin{eqnarray}
\nonumber c = \begin{bmatrix} e^{\,j \theta_{c(1)}}&0&0&0\\
0&e^{\,j\theta_{c(2)}}&0&0
\\  0&0&\ddots &0 \\ 0&0&0&e^{\,j\theta_{c(N)}}
\end{bmatrix} \cdot\begin{bmatrix} |c(1)|\\ |c(2)|\\ \vdots \\ |c(N)|
\end{bmatrix},\\
\end{eqnarray}
\noindent which could be directly fused into the measurement matrix
$\Psi$, thus transforming our model from $\acute{y}=\Psi
c+\acute{z}$ to $\acute{y}=\Psi\Theta_{c} |c|+\acute{z}$ where
\begin{eqnarray}
\nonumber \Psi\Theta_{c}=\begin{bmatrix} |&|&&|\\
e^{\,j\theta_{c(1)}}\psi_{1}&e^{\,j\theta_{c(2)}}\psi_{2}&\dots&e^{\,j\theta_{c(N)}}\psi_{N}
\\|&|&&| \end{bmatrix}
\end{eqnarray}
\noindent has now realigned the phases of the coefficients sought
and reduced the problem to estimating a real sparse vector, with
only the locations and magnitudes of the nonzero coefficients of $c$
to be found. In the case of digital clipping, we can then force the
magnitudes to the nearest alphabets as well.  In any case, with
$\Theta_{c}$ unknown prior to CS, we will instead use
$\Theta_{\hat{\bar{x}}}-2\pi \textbf{I}_{N\times N}$ to augment the
CS algorithm. This could be done in two ways:
\begin{enumerate}
\item \textbf{Sense then Rotate (StR)}: Use the standard CS or weighted CS algorithms used so far to
regain $\hat{c}^{\,\textmd{NoPA}}=\arg_{c\,\in
\textbf{C}^{N}}\min\{\|\acute{y}-\Psi c\|_{2}^{2}+\lambda
\|c\|_{1}\}$ where PA stands for \textit{Phase Augmentation},
extract the locations and magnitudes of the nonzero coefficients
from $\hat{c}$, and then rotate them according to the corresponding
estimated directions in $\hat{\bar{x}}$. i.e.
\begin{eqnarray}
\left\{\hat{c}^{\,\textmd{StR}}(i)\right\}_{i\in
\hat{\mathcal{I}}_{c}}=\left\{|\,\hat{c}^{\,\textmd{NoPA}}(i)|\,e^{j\left(\theta_{\hat{\bar{x}}(i)}-2\pi\right)}
\right\}_{i\in \hat{\mathcal{I}}_{c}} \label{StR}
\end{eqnarray}

\item \textbf{Rotate then Sense (RtS)}: In this case supply the CS
algorithm with the phase information from $\hat{\bar{x}}$ as
described above. This rotation prior to compressive sensing recasts
the problem as an estimation of a real vector with $2m$ real
observations. Defining $\tilde{\Psi}_{c}=\Psi\Theta_{c}$, we're left
with the following model
\begin{eqnarray}
\tilde{y}=\left[\begin{array}{c}
  \Re\acute{y} \\
  \Im\acute{y}
\end{array}\right] = \left[\begin{array}{c}
                \Re\tilde{\Psi}_{c} \\
                \Im\tilde{\Psi}_{c}
              \end{array}\right]\cdot |c| + \left[\begin{array}{c}
                \Re\acute{z} \\
                \Im\acute{z}
              \end{array}\right]
\end{eqnarray}
for which we use the following program to recover $c$
\begin{eqnarray}
\hat{c}^{\,\textmd{RtS}}=\arg_{|\,c\,|\in\,
\textbf{R}^{N}}\min\left\{\|\tilde{y}-\tilde{\Psi}_{\hat{\bar{x}}}
\,|\,c\,|\,\|_{2}^{2}+\lambda \|c\|_{1}\right\} \label{RtS}
\end{eqnarray}
where
$\tilde{\Psi}_{\hat{\bar{x}}}=\Psi\left(\Theta_{\hat{\bar{x}}}-2\pi
\textbf{I}_{N\times N}\right)$. Notice that, similar to (\ref{StR})
one could also replace the phases of $\hat{c}^{\,\textmd{RtS}}$ with
$\left\{e^{j\left(\theta_{\hat{\bar{x}}(i)}-2\pi\right)}\right\}_{i\in
\hat{\mathcal{I}}_{c}}$ after (\ref{RtS}) but we have not observed
any significant improvement in doing so.
\end{enumerate}

%\subsection{Iterative Data-Aided Estimation}
%
%Even when weighted CS was used to improve the peak-reducing signal
%based on information from the data, the estimation of the
%transmitted (clipped) OFDM signal and the peak-reducing signal were
%assumed to be independent throughout this work. Strictly speaking we
%always assumed we had no frequency information about $c$ beyond the
%$m$ observations, since we were unable to distinguish its
%proportions on the other $N-m$ tones from those of $x$. This is not
%strictly true if we could iteratively gain more and more confidence
%in these proportions, which should be possible if the following
%scheme is pursued.
%
%Once we have an estimate $\hat{c}^{cs}$ of the peak-reducing signal
%from CS, we add it back to $\hat{\bar{x}}$ to regain an overall
%estimate of $\hat{x}$ which we can then transfer back to the
%frequency domain and perform QAM detection. This gives us a clean
%estimate of the data vector $\hat{\check{d}}$ after detection, and
%we can assume that the difference between the spectral
%representation of $F\hat{x}$ before and after QAM detection is a
%rough representation of the frequency content of $c$ (i.e. the
%spectral distortion from clipping). In this case we no longer have
%$m$ noisy observations of $c$ but rather $N$, and we can use a more
%reliable method to estimate $c$ given the increased number of
%observations. This refined estimate of $c$ can now be added back to
%$\hat{\bar{x}}$, frequency transformed, and QAM detected again. The
%method should provide a better overall estimate of the data provided
%the initial estimates are good.

\section{Bayesian Estimation of Sparse Clipping Signals} \label{bcs}

To take into account the statistical information at hand, we could
simply modify the dual stage estimate in (\ref{ls cs}) to a linear
minimum mean-square (LMMSE) estimate of the amplitudes $v_{c}$
conditioned on the support estimate $\hat{\mathcal{I}}^{cs}_{c}$
\begin{eqnarray}
\nonumber
\hat{v}_{c}^{lmmse|\hat{\mathcal{I}}^{cs}_{c}}=\sigma_{v_{c}}^{2}
\hat{\Phi}^{H}\left(\sigma_{v_{c}}^{2} \hat{\Phi}
\hat{\Phi}^{H}+\sigma_{z}^{2}I\right)^{-1}
\left(\acute{y}-\hat{\Phi}E_{v_{c}}\right). \label{mmse cs}
\end{eqnarray}

%\begin{equation}
%\hat{v}^{lmmse}_{c}=\sigma_{v_{c}}^{2}
%\hat{\Phi}^{H}\left(\sigma_{v_{c}}^{2} \hat{\Phi}
%\hat{\Phi}^{H}+\sigma_{z}^{2}I_{m}\right)^{-1}\left(\acute{y}-\hat{\Phi}E[v_{c}]\right)+E[v_{c}].\label{mmse
%cs}
%\end{equation}

%\begin{figure}%[htb]
%\centering
%\includegraphics[width=2.5in]{two_sided_pdfs_c_PS.eps}
%\caption{Probability Density Function of $c$ for Peak Suppression as
%a Function of $\gamma$} \label{pdf_c_ps_illustration}
%\end{center} \end{figure}

This should clearly improve upon the least square estimate (\ref{ls
cs}) in case the distribution of $v_{c}$ is Gaussian, but will not
be able to invoke any statistical information into the support
estimate. Using a Maximum a Posteriori (MAP) estimate
$\hat{c}=\arg\max P(\acute{y}|c)P(c)$ generally leads to non-convex
optimization problems in sparse models, and we refer instead to an
MMSE estimate. First define $J^{|\mathcal{I}|}$ as the Hamming
vector of length $N$ and Hamming weight $|\mathcal{I}|$ with active
coefficients according to the support set $\mathcal{I}$. Then
marginalizing on all such possible vectors we obtain
\begin{eqnarray}
\nonumber \hat{c}^{\,\textmd{MMSE}}&=&E\left[\,c\,|\,\acute{y}\,\right]\\
&=&\sum_{i=1}^{2^{N}}E\left[\,c\,|\;
\acute{y},J_{i}\right]P(\acute{y}|J_{i})P(J_{i}) \label{MMSE2}
\end{eqnarray}
\noindent with dropping off $P(\acute{y})$ in (\ref{MMSE2}) due to
its independence of $i$. The estimate is a weighted sum of
conditional expectations, and the formal (exact) approach requires
computing $2^{N}$ terms which is a formidable task for large $N$. To
limit the search space, the key is to truncate the summation index
to a much smaller subset of support vectors $\textbf{J}^{*}$. As
such, the weights $\{P(J_{k}|\acute{y})\}_{k\in J^{*}}$ will not sum
up to unity, and we will need to mitigate this by normalizing the
truncated weighted sum by the sum of weights $\mathcal{W}=\sum_{k\in
\textbf{J}^{*}}P(\acute{y}|J_{k})P(J_{k})$, hence reducing
(\ref{MMSE2}) to
\begin{eqnarray}
\hat{c}^{\,\textmd{MMSE}}&\approx&\frac{1}{\mathcal{W}}\sum_{k\in
\textbf{J}^{*}}E\left[\,c\,|\;
\acute{y},J_{k}\right]P(\acute{y}|J_{k})P(J_{k}). \label{approx
MMSE}
\end{eqnarray}

In effect, estimating $c$ in an MMSE criterion boils down to
appropriately selecting $\textbf{J}^{*}$ and evaluating the terms
$P(J_{k})$, $P(\acute{y}|J_{k})$, and $E\left[\,c\,|\;
\acute{y},J_{k}\right], \forall J_{k} \in \textbf{J}^{*}$, which are
in increasing complexity in the order we've just mentioned.

\begin{figure}%[htb]
\centering
\includegraphics[width=2.5in]{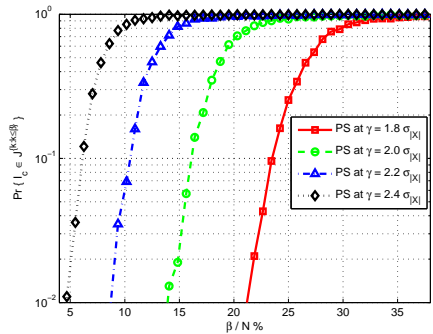}
\caption{Probability of support index set $\mathcal{I}_{c}$ being
completely included in the first $\beta/N \%$ of
$\arg\{w^{\downarrow}\}$} \label{Probability_of_inclusion}
\end{figure}

When using peak suppression to $\gamma$, the receiver is given a
vague picture of where clipping has occurred based on the affinity
of $\hat{\bar{x}}$ to $\gamma$. Consequently, by sorting the
magnitudes of the weighting vector $w^{\downarrow}$ in
(\ref{ps_distance}) in ascending order, the probability of the true
support coinciding with the first $\beta$ elements in
$\arg\{w^{\downarrow}\}$ will increase rapidly with $\beta$. Fig.
\ref{Probability_of_inclusion} shows a Monte Carlo simulation of
this probability at different clipping thresholds. For instance,
this implies that given a clipping threshold of
$\gamma=2\sigma_{|X|}$, one could exclude $70\%$ of the $N$ indices
as having too low a probability of corresponding to a clipping
location, thus reducing the possible candidates from $2^{N}$ to
$2^{\beta}$ Hamming vectors.

Given this reduced set $\textbf{J}^{\{k:k\leq\beta\}}$ of vectors,
we adopt a search over it by latching a vector of unity Hamming
weight based on (\ref{approx MMSE}), and then proceed in a greedy
fashion similar to Larsson \cite{Larsson} and Schniter
\cite{Schniter1,Schniter2} until a maximum sparsity level $s^{max}$
is reached. This will preserve the quality of the greedy estimate
using Fast Bayesian Matching Pursuit (FBMP) in \cite{Schniter1}
while reducing the number of executions of (\ref{approx MMSE}) by
\begin{eqnarray}
\nonumber 100\left( 1-\frac{\beta(1+\rho \cdot s^{max})-\frac{\rho
\cdot s^{max}(s^{max}+1)}{2}}{N(1+\rho \cdot s^{max})-\frac{\rho
\cdot s^{max}(s^{max}+1)}{2}} \right)\%
\end{eqnarray}
where $\rho$ is the number of tested candidates for each Hamming
weight. This would correspond to a reduction of $60-80\%$ of
executions with our practical parameters, and we will henceforth
refer to this procedure as $\beta$-FBMP.

\section{Performance Analysis and Simulations}

%Capacity alone is not a sufficient performance parameter. In other
%words, we cannot justify reserving tones based on the amount of SER
%reduction alone, since our primary objective is reducing the dynamic
%range and hence the cost of power amplifiers while trying to keep
%capacity unharmed.

For our simulation purposes we considered an OFDM signal of $N=256$
subcarriers of which $m=0.2N$ are randomly dispersed measurement
tones. The data coefficients were generated from a QAM constellation
of size $M=32$. The Rayleigh fading channel model was of 32 taps,
operating at a $30$ dB SNR environment. The performance parameters
we considered were the SER, the relative temporal complexity, the
PAPR reduction ability, and the capacity.

Our primary objective was to test the SER variation with the
clipping threshold $\gamma$ for a clipped OFDM signal that used our
different adaptations of CS algorithms and clipping techniques.
Observed as a variable, the clipping threshold in particular is of
central importance due to its critical effect on both CS generic
performance and the PAPR reduction. Decreasing $\gamma$
significantly reduces the PAPR but also implies a nonlinear increase
in the average sparsity level that the estimation algorithms must
tolerate. It also has a positive counter effect on CS performance as
well since it increases the CNR, making the overall behavior of
SER($\gamma$) difficult to predict.

Furthermore, when testing the precise performance of an algorithm we
used the Normalized Mean Square Error
\begin{eqnarray}
\nonumber
\textmd{NMSE}&=&E\left[\frac{(c-\hat{c})^{2}}{\|c\|_{2}^{2}}\right]
\end{eqnarray}
%\begin{eqnarray}
%\nonumber \textmd{NMSE}&=&E\left[\frac{(c-\hat{c})^{2}}{\|c\|_{2}^{2}}\right]\\
%&=&\frac{1}{N}\sum_{i=1}^{N}\frac{|c(i)-\hat{c}(i)|^{2}}{|c(i)|^{2}}
%\end{eqnarray}
to ensure that error decrease was not simply due to a decrease in
the number of estimated variables.

\begin{figure}
\centering
\includegraphics[width=2.5in]{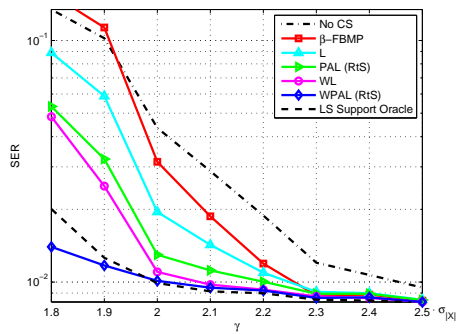}
\caption{SER of PS vs $\gamma$} \label{SER_gamma_PS}
\end{figure}

\begin{figure}
\centering \includegraphics[width=2.5in]{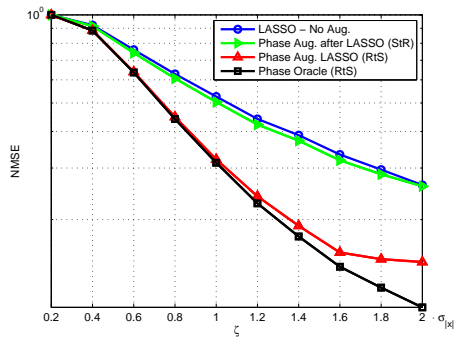}
\caption{NMSE of Digital Clipper estimate as a function of the
coefficient magnitude $\zeta$} \label{DC_NMSE_vs_zeta}
\end{figure}

\begin{figure}%[htb]
\centering
\includegraphics[width=2.5in]{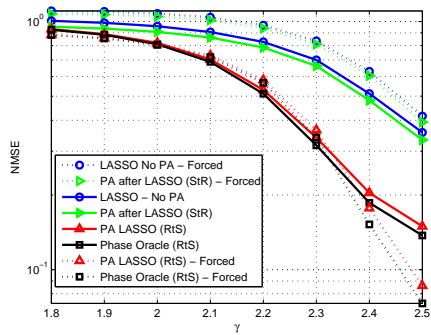}
\caption{NMSE of Digital Clipper Estimate as a function of the
clipping threshold $\gamma$} \label{DC_NMSE_vs_gamma}
\end{figure}

\begin{figure}%[htb]
\centering
\includegraphics[width=2.5in]{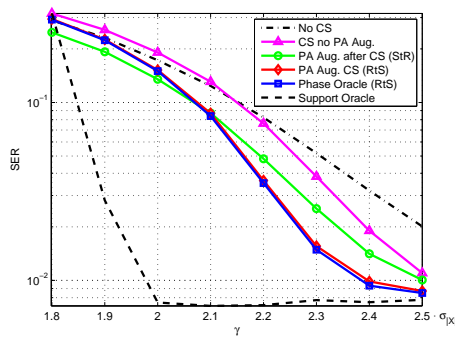}
\caption{SER of Digital Clipping with $\zeta=0.8\,\sigma_{|X|}$ vs
$\gamma$} \label{DC_SER_vs_gamma}
\end{figure}

Fig. \ref{SER_gamma_PS} shows the SER for Peak Suppressing clippers
in \ref{PS} after QAM decoding
$(\textbf{F}\textbf{S}_{x})^{\dagger}(\hat{\bar{x}}^{ls}+\hat{c}^{(ps)})$
as the clipping threshold is varied. The methods tested were the
reduced search space greedy method ($\beta$-FBMP), the LASSO, the
Phase-Augmented LASSO (PAL) using (\ref{RtS}), the data-based
Weighted LASSO (WL), and the Weighted Phase-Augmented LASSO (WPAL).
These were compared against two performance bounds: the lower bound
of not estimating $c$, and the upper bound of an oracle receiver
that knows the support $\mathcal{I}_{c}$, and simply uses least
squares to estimate the coefficients' amplitudes. Interestingly,
combining the support and phase augmentation techniques into the
LASSO enables it to perform very close to the support oracle, and
even beat it at low clipping thresholds where $s>0.55\,m$ since it
has additional information regarding the coefficients' phases.
Furthermore, weighting alone is more effective then
phase-augmentation, although both significantly improve the
performance of the LASSO.

To see the effect of varying the magnitude of active coefficients in
digital clipping of section \ref{dc} we plotted the NMSE vs $\zeta$
in Fig. \ref{DC_NMSE_vs_zeta}. This avoids a biased evaluation due
to increased CNR with $\zeta$. The results imply that embedding the
phase information into the LASSO in (\ref{RtS}) is much more
effective than rotating the estimate after compressed sensing in
(\ref{StR}). It also shows that the former method is considerably
close to a phase oracle that uses the same technique for practical
values of $\zeta$ relative to $\sigma_{|X|}$. However, as expected
they eventually deviate as we increase $\zeta$ since this
corresponds to decreasing the SNR and hence the accuracy of the
phase information induced from the data vector estimate
$\theta_{\hat{\bar{x}}}$. Fig. \ref{DC_NMSE_vs_gamma} implies that
forcing the magnitudes of the estimates in (\ref{StR}) and
(\ref{RtS}) is generally ineffective except in the very sparse cases
for the former. The overall result on the SER is portrayed in Fig.
\ref{DC_SER_vs_gamma} at a fixed $\zeta=0.8\,\sigma_{|X|}$.

%\begin{figure}%[htb]
%\centering
%\includegraphics[width=2.5in]{PS_and_DC_vs_gamma_at_same_CNR_500c.eps} \caption{SER of PS
%and DC compared at the same CNR vs $\gamma$}
%\label{PS_and_DC_vs_gamma_at_same_CNR}
%\end{figure}

%In pursuing fairness when comparing the performance of digital
%clipping and peak suppression, we simulated them at an equivalent
%CNR by regulating $\zeta=f(\gamma)$ according to
%(\ref{CNR_equivalence}) at each level of the simulation versus the
%clipping threshold. According to Fig.
%\ref{PS_and_DC_vs_gamma_at_same_CNR}, Peak Suppression considerably
%outperforms Digital Clipping, which at such low magnitudes only
%begins to outperform the lower bound at $\gamma > 2.24
%\,\sigma_{|X|}$.

%\begin{figure}%[htb]
%\centering
%\epsfxsize = 3.0 true in \epsfbox{DC_SER_vs_gamma_1000_forced.eps}
%\caption{SER of DC vs $\gamma$} \label{DC_SER_vs_gamma}
%\end{figure}

\begin{figure}%[htb]
\centering
\includegraphics[width=2.5in]{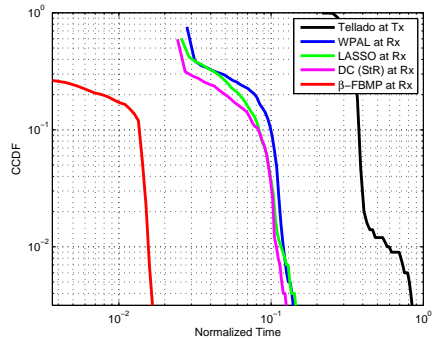}
\caption{CCDF of execution time normalized by maximum value}
\label{temporal_complexity}
\end{figure}

Complexity-wise, we neglect mentioning implementation and orders of
complexity since they match those of standard algorithms we've built
on and that are well documented in the CS literature (e.g.
\cite{Tropp3,Schniter1,Boyd2}). Instead we investigate the practical
aspect of the relative time required to execute the major techniques
proposed in the paper compared to Tellado's primary tone-reservation
algorithm using the same generic CVX software
\cite{Boyd}.\footnote{With the only exception being Schniter's
Greedy algorithm when evaluating $\beta$-FBMP.} As such we collected
the random execution times for $2000$ runs of each, normalized them
by the maximum execution time among all, and plotted their CCDF.
Fig. \ref{temporal_complexity} depicts the results. Roughly
speaking, the methods stemming from the LASSO required less then $12
\%$ of the time required to execute Tellado's primary QCQP algorithm
on average, while the $\beta$-FBMP required less than $2\%$ of the
time.

%\begin{figure}%[htb]
%\begin{center}
%\epsfxsize = 3.0 true in
%\epsfbox{temporal_complexity_CCDF_loglog_dashed.eps}
%\caption{Probability Density Function of $c$ for Peak Suppression as
%a Function of $\gamma$} \label{pdf_c_ps_illustration}
%\end{center} \end{figure}

\begin{figure}%[htb]
\centering
\includegraphics[width=2.5in]{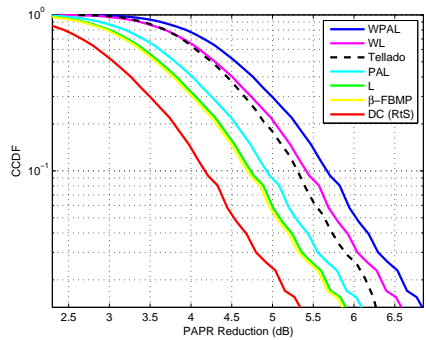}\caption{CCDF
of PAPR Reduction (dB)} \label{CCDF_PAPR_Reduction}
\end{figure}

A major advantage of clipping to a fixed threshold is that, unlike
tone-reservation methods such as \cite{Tellado,Kashin} the dynamic
range, maximum power, and PAPR of the transmitted signal are fixed.
The distribution of PAPR reduction,
$10\log_{10}\left(\frac{P_{max}}{\gamma^{2}}\right)$, would simply
follow from the distribution of the maximum squared coefficient in
$x$ (refer to \cite{Imai,Bahai,Wei} for relevant analysis) which we
plot in Fig. \ref{CCDF_PAPR_Reduction}. The fixed maximum power
followed from the clipping threshold that corresponded to a SER of
$10^{-2}$ for the different techniques in this work.

\begin{table}[ht]
\renewcommand{\arraystretch}{1.3}
\caption{Summary of Results} % title of Table
\centering
\begin{tabular}{|c|c|c|c|}
  \hline
   & Tolerable $\gamma$ & Avg. PAPR Red. (dB) & $\%$ Exec. Time \\
  \hline
  DC (RtS) & 2.40 $\cdot \,\sigma_{|X|}$ & 3.19 & $11.06 \%$ \\
  $\beta$-FBMP & 2.26 $\cdot \,\sigma_{|X|}$ & 3.71 & $1.6 \%$ \\
  LASSO & 2.25 $\cdot \,\sigma_{|X|}$ & 3.75 & $12.3 \%$ \\
  WPAL & 2.02 $\cdot \,\sigma_{|X|}$ & 4.68 & $13.9 \%$ \\
  Tellado & - & 4.37 & $100 \%$ \\
  \hline
\end{tabular}
\label{table:summary} % is used to refer this table in the text
\end{table}

The most fundamental parameter of interest given a desired clipping
threshold is the channel capacity \cite{Tellado,Capacity_clipped_1}
\begin{eqnarray}
\nonumber C=\sum_{k=1}^{N}
\log_{2}\left(1+\frac{|D(k,k)|^{2}\sigma_{x(k)}^{2}}{\sigma_{z(k)}^{2}}\right),
\end{eqnarray}
and we will thus consider two systems. The first system
$\mathcal{S}_{1}$ clips all coefficients above $\gamma$ and does not
reserve tones to estimate the clipping signal $c$, resulting in a
higher clipping noise over all $N$ tones while retaining all of them
for data transmission. The second system $\mathcal{S}_{2}$ reserves
$m$ tones to estimate $c$, thus reducing the SER degradation while
also reducing the data tones by $m$.

The justification then depends very much on the variances of the
clipping noise $\{\sigma_{c}^{2}(k;\gamma)\}_{k\in \Omega_{d}}$ with
and without estimation at the receiver. Furthermore, if the
threshold $\gamma$ is sufficiently low relative to $\sigma_{|X|}$
(e.g. $E\left[\|c\|_{0}\,;\gamma\right]=10\%$ of $N$), the clipping
noise on each tone will be the result of a reasonably large
summation of scaled coefficients of $c$ in the time domain, and so
will the distribution of the priors in (\ref{pdf_c_ps}) converge to
a Gaussian. With this theoretical justification aided by extensive
simulations, we will assume for simplicity that the distortion on
each carrier follows a Gaussian with a common variance
$\sigma_{c}^{2}$. However, caution must be taken when comparing this
parameter for the two systems. The reason is that $\mathcal{S}_{1}$
has more data energy than $\mathcal{S}_{2}$ by using all $N$ tones,
and will thus have a higher distortion variance at the same clipping
level $\gamma$, i.e.
$\sigma_{c\,\mid|\Omega_{d}|=N}^{2}>\sigma_{c\,\mid|\Omega_{d}|=N-m}^{2}$.
Consequently, the capacity of the first system (after dropping the
tone index) will be
\begin{eqnarray}
C_{1}=N\log_{2}\left(1+\frac{|D|^{2}\sigma_{x\,\mid|\Omega_{d}|=N}^{2}}{|D|^{2}\sigma_{c\,\mid|\Omega_{d}|=N}^{2}+\sigma_{z}^{2}}\right)
\label{C1}
\end{eqnarray}
while the capacity of the second will be
\begin{eqnarray}
C_{2}=(N-m)\log_{2}\left(1+\frac{|D|^{2}\sigma_{x\,\mid|\Omega_{d}|=N-m}^{2}}{|D|^{2}\sigma_{\left(c-\hat{c}\right)\mid|\Omega_{d}|=N-m}^{2}+\sigma_{z}^{2}}\right)
\label{C2}
\end{eqnarray}
The use of reserved tones for CS is then justified if $C_{2}>C_{1}$,
i.e. when
\begin{eqnarray}
\sigma_{\left(c-\hat{c}\right)\mid|\Omega_{d}|=N-m}^{2}<\frac{\sigma_{x\,\mid|\Omega_{d}|=N-m}^{2}}{\left[1+\frac{|D|^{2}\sigma_{x\,\mid|\Omega_{d}|=N}^{2}}{|D|^{2}\sigma_{c\,\mid|\Omega_{d}|=N}^{2}+\sigma_{z}^{2}}\right]^{\frac{N}{N-m}}-1}-\frac{\sigma_{z}^{2}}{|D|^{2}}
\label{C_condition}
\end{eqnarray}

\begin{figure}%[htb]
\centering
\includegraphics[width=2.5in]{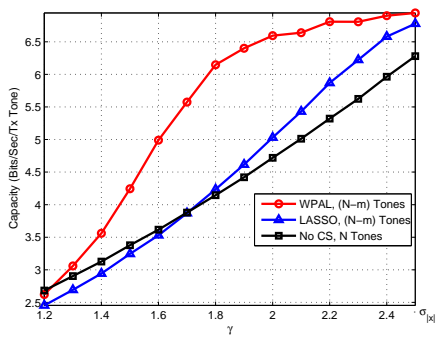}
\caption{Capacity per transmitted tone at different clipping
thresholds} \label{capacity_vs_gamma}
\end{figure}

It would be very interesting to observe how this parameter behaves
as a function of the clipping threshold $\gamma$ as both the
distortion $\sigma_{c}^{2}$ and the quality of the estimate
$\hat{\sigma}_{c}^{2}$ nonlinearly counteract each other. Fig.
\ref{capacity_vs_gamma} shows such results upon $1000$ runs at each
$\gamma$ for estimating $\sigma_{c}^{2}$ and
$\sigma_{\left(c-\hat{c}\right)}^{2}$. The results show that by
reserving $20\%$ of the tones for data-based weighted and
phase-augmented LASSO the capacity of such a system can
significantly outperform the naive system which uses all the tones
for data transmission. What's more, the capacity associated with
this technique behaves in a convex fashion so that by reducing the
capacity by less then 1 bit per second per transmitted tone, the
clipping threshold can be dramatically reduced from
$\gamma=2.5\,\sigma_{|X|}$ to $\gamma=2\,\sigma_{|X|}$. Unlike the
semi-linear relation of $\mathcal{S}_{1}$ with $\gamma$, such
behavior offers a very tempting compromise between capacity and
peak-reduction. Using the typical LASSO at such conditions is
effective at clipping thresholds reaching as low as
$1.9\,\sigma_{|X|}$ which is impressive.

\begin{figure}%[htb]
\centering
\includegraphics[width=2.5in]{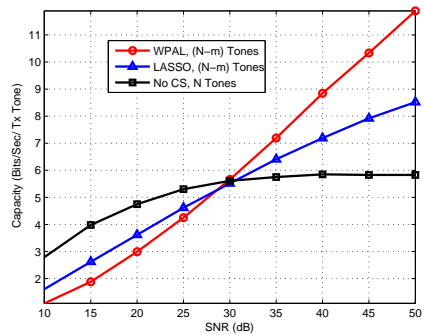}
\caption{Capacity per transmitted tone vs SNR}
\label{capacity_vs_SNR}
\end{figure}

Fig. \ref{capacity_vs_SNR} implies that increasing the SNR is much
more rewarding for $\mathcal{S}_{2}$ compared to $\mathcal{S}_{1}$
which we test at a fixed clipping threshold of $2.3\,\sigma_{|X|}$.
The reason is that eliminating $\sigma_{z}^{2}$ has no effect on
$\sigma_{c}^{2}$ and the capacity of the naive system saturates
after an SNR of 35 dB. On the other hand, decreasing the noise level
improves the CS estimate and hence has a dual effect in increasing
the capacity, leading to the semi-linear relation with the SNR.
\section{Conclusion}

In this work we have established the new general concept of clipping
mitigation (and hence PAPR reduction) in OFDM using compressive
sensing techniques. The general framework stressed the use of
reserved subcarriers to compressively estimate the locations and
amplitudes of the clipped portions of a transmitted OFDM signal at
the receiver, instead of using them at the transmitter as a spectral
support for optimized peak reducing signals in the time domain.
Consequently, the method interchanges the \emph{stage} at which
signal processing complexity is required compared to the previous
techniques, hence introducing a real solution to communication
systems that use OFDM signals at the physical layer and require
minimal complexity at the transmitter.

The other major contribution is demonstrating how by a marginal
increase in complexity one can augment the standard $\ell_{1}$
minimization of CS by extracting information regarding clipping
locations, magnitudes, and phases from the data, and hence enable
the system to estimate sparse clippers far beyond the recoverability
conditions of CS (e.g. sparsity levels above $55\%$ of $m$). Such
augmentation was shown to significantly boost the overall system's
capacity at low clipping thresholds and thus suggests a very
appealing compromise between capacity and peak-reduction.
\ifCLASSOPTIONcaptionsoff
  \newpage
\fi




\bibliographystyle{IEEEtran}

%\bibliography{IEEEabrv,safadi_jnl}


%\bibliography{safadi_jnl}{}

\bibliographystyle{plain}

\newpage

\end{document}